\documentstyle[11pt]{article}
\begin{document}
\begin{center}
{\Large \bf Sudakov Form Factors in Leptoproduction of Vector Mesons}\\
\bigskip
{\large I.M. Dremin}\\
\bigskip
{\normalsize Lebedev Physical Institute, 117924 Moscow, Russia}\\
\end{center}
\begin{abstract}
It is shown that Sudakov form factors for a colour dipole in a QCD-inspired 
model of leptoproduction of vector mesons reduce the value of the cross section
of the process by an order of magnitude. They suppress the large size
quark-antiquark pairs and unequal sharing of energy among the components
of the dipole.
Some freedom in the choice of the model parameters is also discussed.
\end{abstract}
The idea of inelastic diffraction goes back to the papers of I.Ya. Pomeranchuk
and E.L. Feinberg \cite{pf}. The particular example is provided by the
diffraction dissociation of a virtual photon to any vector meson on a nuclear
target. Recently,
the processes of diffractive leptoproduction of vector mesons in interactions 
of high energy electrons and muons with nucleons and nuclei were 
intensively investigated both experimentally and theoretically. The most 
popular approach to the problem of their description is provided by a model
inspired by ideas of quantum chromodynamics (QCD) where the exchanged photon
with the four-momentum squared (the virtuality) $Q^2$ is treated as a 
quark-antiquark pair which interacts with a target nucleon by the two-gluon
exchange \cite{1,2,3,4}. Correspondingly, in the simplest approximation the 
diffractive cross section of "elastic" leptoproduction of vector mesons is 
defined by
a product of wave functions of the virtual photon and of the vector meson 
with the amplitude of elastic scattering of the quark-antiquark dipole off 
the nucleon. 

This factorization model becomes valid at high energies because, due to the 
Lorentz transformation, those longitudinal distances at which the photon
separates as the colour dipole (and, later, this dipole transforms into the
vector meson) exceed strongly the range of distances of the intermediate 
interaction of the dipole with the target nucleon. That is why the whole 
process can be treated as if separated into three independent stages. First,
long before its contact with the target nucleon, the virtual photon forms
the quark-antiquark component of its wave function. At the second stage, 
the scattering of this colour dipole on the target takes place during the 
rather short time when the transverse size of the dipole does not change much.
This subprocess used to be described by the two-gluon exchange at high energies.
Finally, at the third stage, the vector meson is formed by the quark-antiquark
pair, and it takes long time comparable with the initial stage of formation
of the quark-antiquark pair from the virtual photon.

However, after writing down the leptoproduction matrix element as a product of 
the wave functions and of the dipole cross section (inserted in place of the 
imaginary part of the forward scattering amplitude) one has to solve a 
problem of ascribing definite expressions to all the terms in the product.
Usually, the cross section is represented either by the formulae obtained in
the leading-logarithmic approximation of the perturbative QCD (and replaced 
sometimes by some suitable fits) or by the expressions rooted in the reggeon
approach. The wave functions of vector mesons are described by different 
semiphenomenological forms. 

The band of undefiniteness due to the assumptions 
of such a kind has been studied in Ref. \cite{3}. It has been shown that they 
give rise to the factors about 2 or 3 in the forward differential cross section
of the leptoproduction, and consequently to the same factors in its total cross 
section which differs from above characteristics by the diffraction cone slope 
only. 

In the paper \cite{4} it has been demonstrated that the similar factors
can appear due to the additional rescattering of the quark-antiquark pair on
the nucleon giving rise to the shadowing corrections to the initial expressions.
The rescattering of the colour dipole was considered in Ref. \cite{4} as mediated 
by the exchange of the ladder-type graphs taking into account the gluon 
distribution within the nucleon. All terms of that kind have been summed up.
Suppression effects increasing with $Q^2$ have been claimed.

Corrections due to quark Fermi motion within produced vector mesons were
considered in the paper \cite{fks}.

However, up to now nobody took into account that due to the exclusive nature of 
the process of leptoproduction of vector mesons one must consider its 
suppression by Sudakov form factors. Physical origin of this effect consists 
in the requirement of prohibition of any emission of additional "free" gluons
(gluon jets) during the whole process of the conversion of the virtual 
photon into the vector meson. Otherwise, we would have to deal with inelastic 
processes of a more general type. In other words it can be formulated as the 
requirement to deal with those components of the quark-antiquark pair plus
one or several gluons which form the Fock column of the bound state and have
no "free" gluon jets since the last ones would correspond to the processes
not considered here. The reason for such a neglect of this effect is rooted
probably in the necessity to use expressions for Sudakov form factors of the
quark-antiquark dipoles (yet not very widely known) instead of the famous 
double-logarithmic formulae for form factors of single quarks. However the 
corresponding formulae for the colour dipole were obtained rather long time
ago in the paper \cite{5} where they were used for exclusive processes of 
elastic meson-meson and baryon-baryon scattering at high energies.

The aim of the present paper is to reveal the role of Sudakov form factors of
quark-antiquark dipoles in suppression of the cross section of leptoproduction
of vector mesons. We shall show that the corresponding damping factors are
larger than other corrections discussed above. Here we shall consider the 
simplest analytical estimates leaving detailed computer calculations for a
more complete subsequent publication.

The ratio $R$ of the cross section values with and without Sudakov form 
factors in the model under consideration is given by
\begin{equation}
R=\left[ \frac {\int dz \int d^{2}\rho \psi _{\gamma ^*}\sigma \psi _{V} e^{-S}}
{\int dz \int d^{2}\rho \psi _{\gamma ^*}\sigma \psi _{V}}\right] ^2.    \label{1}
\end{equation}
Here $\psi _{\gamma ^*}$ and $\psi _{V}$ are the wave functions of the virtual
photon and of any possible vector meson, correspondingly, $\sigma $ is the cross section 
of interaction of the colour dipole of the transverse size $\rho $ with the 
target nucleon, $S$ describes Sudakov suppression for the colour dipole, $z$
is the share of energy of the virtual photon carried by the quark (the
antiquark share of energy is equal to $1-z$, correspondingly). The derivation 
of the formulae for leptoproduction matrix elements is given in papers 
\cite{1,2,3},
and we do not discuss it here. Let us just mention that such a representation
of matrix elements becomes possible because the dipole size $\rho $ is 
practically constant during the intermediate stage of scattering of the dipole 
on the nucleon as discussed above.

Since we are interested here just in the ratio $R$ given by (1) and not in the
absolute value of the leptoproduction cross section, we shall describe general dependences
of the wave functions $\psi $ and the scattering cross section $\sigma $ on
$z$ and $\rho $ leaving aside all constant factors cancelled out in the ratio 
$R$. Thus the wave function of the virtual (longitudinally polarized) photon
is proportional to $z(1-z)K_{0}(\epsilon \rho )$, where $K_0$ is the 
modified Bessel function of the second kind,
\begin{equation}
\epsilon ^2 = z(1-z)Q^2 +m^2,     \label{2}
\end{equation}
$m$ is the quark mass. The wave function of the vector meson has either 
similar dependence on $z$ or acquires another factor $(1-2z)^2$. The first 
case corresponds to the asymptotical form of the distribution while the 
second one to the form with suppressed contribution of the symmetrical
shares of energy suggested for pions in the paper \cite{6} when considering
the QCD sum rules. The cross section of the interaction of the colour dipole 
with the nucleon is usually described either by an expression relating it to 
the gluon structure function so that $\sigma \propto \rho ^2 xG(x,Q^2)$, where
one uses any of the successful parametrizations of experimental data obtained 
at HERA for the structure function $G(x,Q^2)$, or by the semiphenomenological 
formula \cite{1} $\sigma \propto \rho^2 \log (1+R_{N}^{2}/\rho ^2)$, where
$R_N$ is the nucleon radius, which takes into account that the strong 
interaction cross section is bounded by the particle size.

The typical common feature of these formulae in both approaches is the 
smallness of the dipole cross section  for small size dipoles due to the 
factor $\rho ^2$. It is called the colour transparency effect. Its physical 
origin is well known from its analogue in quantum electrodynamics which
is called Chudakov effect and ascribed to the mutual screening of the 
opposite sign electric charges of the components of an electron-positron pair
at a small space separation between them. It suppresses the role of small size
dipoles and, therefore, increases the effective average 
size of dipoles conributing to the process.

The definite choice of various forms of the wave functions and the dipole
cross section as well as their normalization are discussed at some length in
the cited papers \cite{1,2,3,4}. That is why we shall not discuss them in more
details here and turn now to the Sudakov form factors of the colour dipole.
In the paper \cite{5} all Sudakov effects were organized into the wave function
itself and written down as functions of $z$ and $\rho $ suitable for our 
purposes according to (1). The asymptotical form of Sudakov suppression factors 
has been given in Ref. \cite{5}. Let us consider the first term of it which
provides the dominant contribution at high virtualities. It can be written as
\begin{equation}
S=\frac {c}{2}[\log \frac {zQ}{\Lambda }\log ({\frac {\log \frac {zQ}{\Lambda }}{\log \frac
{1}{\rho \Lambda }}})+\log \frac {(1-z)Q}{\Lambda }\log ({\frac {\log \frac 
{(1-z)Q}{\Lambda }}{\log \frac {1}{\rho \Lambda }}})],    \label{3}
\end{equation}
where the constant $\Lambda $ defines the QCD scale in the running coupling 
constant, $c=32/(33-2n_f)$, $n_f$ denotes the number of active flavours. We
consider the case $n_f=3$, and therefore $c \approx 1.2$. Let us mention that 
the expression (3) reminds closely the earlier obtained formula (15) of the
unpublished preprint \cite{a}.

In Eq. (1), small 
dipoles are damped by the Chudakov effect, i.e., by the factor $\rho ^2$ in
$\sigma $, while large dipoles by the wave functions, i.e., by the factor
$K_{0}(\epsilon \rho )$. Let us note that Sudakov form factors suppress large 
dipoles as well. For wavelengths long compared to $\rho $, radiation is damped 
by destructive interference of the dipole components and $1/\rho $ acts as an 
infrared cutoff. Therefore, radiation forces $\rho $ to smaller sizes. However
the cutoff provided by the Bessel function in the wave functions appears to be
more effective.

Let us show that these statements follow from Eqs. (3) and (1). Sudakov 
suppression in (1) is given by the factor
\begin{equation}
e^{-S}=\log ^{\kappa}\frac {1}{\rho \Lambda }(\log \frac {zQ}{\Lambda })^{-
\frac {c}{2}\log \frac {zQ}{\Lambda }}(\log \frac {(1-z)Q}{\Lambda })^{-
\frac {c}{2}\log \frac {(1-z)Q}{\Lambda }},    \label{4}
\end{equation}
where
\begin{equation}
\kappa =c\log \frac {\epsilon }{\Lambda }    \label{5}
\end{equation}
(quark masses are neglected in $\epsilon $). Large sizes of the dipole are
damped exponentially by the wave function and get additional damping 
Sudakov factor (4) which imposes further requirement that they should not 
exceed the range of nuclear forces $\rho \leq 1/\Lambda $. Therefore to
simplify our model for analytical estimates we can neglect the factors like 
$\log (1+R_{N}^{2}/\rho ^2 )$ in the dipole cross section and deal with its
$\rho ^2$-increase only because dipoles of relatively small sizes are 
important. It is easily seen from Eq. (4) that the value of $\epsilon ^2 \approx
z(1-z)Q^2$ which accounts for conservation laws in the colour dipole properly
sets up the scale of corrections but not $Q^2$ itself.

The effective value of the dipole size and its diminishing due to Sudakov
suppression are estimated qualitatively if one inserts the simplest 
analytical expressions for wave functions and dipole cross section
discussed above. The $\rho $-integral in the numerator of (1) can be 
approximated by
\begin{equation}
I_{\rho }=\epsilon ^{-1/2} \int d\rho \rho ^{5/2} e^{-\epsilon \rho }\log ^
{\kappa }\frac {1}{\rho \Lambda },       \label{6}
\end{equation}
where $K_{0}(\epsilon \rho )$ is replaced by its asymptotics proportional to
$(\epsilon \rho )^{-1/2}e^{-\epsilon \rho }$. Therefore the effective values
of $\rho $ at large virtuality $Q^2$ are given by
\begin{equation}
\rho _{eff}=\frac {\beta }{\epsilon } \approx \frac {2.5-c(1+\frac {\log (2.5-c)}
{\log \frac {Q}{2\Lambda }})}{\epsilon }.    \label{7}
\end{equation}
From Eq. (7) it follows that the asymptotical formula for $K_0$ can be used
at $\rho _{eff}$ with the accuracy better than 10$\%$ that is enough for our 
purposes. Moreover, the asymptotical expression imitates in some way the 
neglected "logarithmic" term of $\sigma $. We would like to stress that the
asymptotical value of $\rho _{eff}$ does not depend on infrared sensitive
value of $\Lambda $ because the effective size of the dipoles is much less
than the radius of nuclear forces.
In the logarithms of the preasymptotical term in (7) we have replaced the $\beta $
value by its asymptotics ($2.5-c$) and inserted $Q/2$ instead of $\epsilon $
that corresponds to symmetrical sharing of energy among components of the 
dipole. To demonstrate the validity of the latest statement, we turn again to 
the formulae (3) or (4) and evaluate Sudakov suppression of symmetrical and 
asymmetrical pairs substituting now the effective value of $\rho $ given by Eq. (7).

For symmetrical pairs with $z=1-z=1/2$ one gets
\begin{equation}
e^{-S}\vert _{symm} \approx \beta ^{-c} (1+O(\frac {1}{\log Q/\Lambda })), \label{8}
\end{equation}
i.e., for small dipole sizes (7), the suppression becomes a numerical one 
instead of parametrical. Therefore, it does not change $Q^2$-dependence in
this approximation.

Near its maximum at $z=1/2$ the Sudakov factor varies quite slowly and is 
approximated by
\begin{equation}
e^{-S}\approx \beta ^{-c}\exp[-\frac {2c(z-1/2)^2}{\log Q/2\Lambda }]. \label{9}
\end{equation}

For extremely asymmetrical pairs with $z=\Lambda /Q$ or $1-z=\Lambda /Q$
the suppression becomes much stronger at large $Q$
\begin{equation}
e^{-S}\vert _{asymm} \approx \beta ^{-c}\left( \frac {\Lambda }{Q}\right) 
^{\frac {c}{2}\log 2}.   \label{10}
\end{equation}
The physical origin of this result is quite clear because a symmetrical pair is 
less inclined to emission of gluons due to stronger colour screening compared
to an asymmetrical pair with tendency to the space separation of its colour 
components. In other words it can be stated if one treats the exchanged gluons 
as probes of the internal structure of the colour dipole. The quark-antiquark 
pair of size $\rho $ can be resolved by the gluons with energy exceeding 
$1/\rho $. Therefore, Sudakov suppression is the stronger the farther one from 
another are the components of the pair and, consequently, the points of 
interaction with the nucleon.

One can evaluate the upper bound of $R$ replacing the integrand in Eq. (6)
by the expansion near its maximum $\rho =\rho _{eff}$  given by (7) and
considering the corresponding Gaussian integral. At large $Q^2$  we have
\begin{equation}
R\leq \left( \frac {\beta }{\beta _0}\right) ^{6}\left( \frac {e}{\beta }\right)
 ^{2c}, \label{11}
\end{equation}
where $\beta _{0}=\beta (c=0)=2.5$. The factors in front of $\beta ^{-2c}$
account for the diminished size of the dipole due to Sudakov suppression.
The Chudakov effect combined with phase space factors is in charge of the high 
power of $\beta $ in (11). The second factor comes from Sudakov form factors
and wave function increase at smaller distances. Therefore, in total it 
results in an asymptotically constant factor. Strong $Q$-dependence of 
Sudakov form factors at constant $\rho $ becomes cancelled due to smallness
of effective $\rho $-values.
No dependence on the strong forces scale $\Lambda $ appears in the
asymptotical expressions again.

 For asymptotical estimates one can insert
\begin{equation}
\beta \rightarrow \beta _{as}=2.5-c\approx 1.3.   \label{12}
\end{equation}
Thus one gets at large $Q^2$
\begin{equation}
R\leq 0.12.     \label{13}
\end{equation}
Therefore Sudakov form factors imposed on Chudakov effect suppress the 
cross section of leptoproduction of vector mesons by an order of magnitude,
at least.
This damping is much stronger than all the corrections previously considered.
Let us stress again the new feature revealed by our analysis that, for dipole 
sizes as small as given by (7), strong cancellations appear in (3), form factors
become constant and do not depend on $Q^2$.

To get more definite conclusions (in particular about $Q^2$ dependence of the 
correction terms) one needs accurate quantitative estimates according to Eq. (1)
taking into account non-leading terms in $S$, different forms of the wave 
functions and the dipole cross section. We have treated the leading term of 
dipole form factors (3). However, the preasymptotical terms should be included
as well for virtualities available in experiment. Besides, it becomes especially
important in view of strong cancellations in $Q^2$-dependence of form factors 
even for large virtualities due to small dipole sizes as demonstrated above.
We intend to do it as soon as possible.

Still, one can get some qualitative conclusions right now. For example, it is 
clear that Sudakov damping is stronger for wave functions with suppressed equal 
sharing of energy like those ones given in Ref. \cite{6}. 

Also, the relative role of Sudakov form factors and of the wave functions is
essential in meson-meson, meson-baryon and baryon-baryon elastic scattering
as considered in Ref. \cite{5}. It can change somewhat 
the conclusions of Ref. \cite{5}
about the saddle point location and the relation between the Landshoff model
and dimensional counting rules. The relative contributions of symmetrical and 
asymmetrical pairs in nuclear shadowing should be reconsidered as well. All
these problems will be treated in a subsequent publication.

To conclude, we have shown that Sudakov suppression of the cross section of 
leptoproduction of vector mesons is quite strong if the term (3) is still
prevailing for small size dipoles (7). The 
general effect of Sudakov form factors is further suppression of large size 
colour dipoles and of unequal sharing of energy among its components apart
from effects due to the wave functions.

\bigskip

{\large  \bf Acknowledgement}\\

\bigskip

I am indebted to Hoi-Lai Yu who first pointed out the paper \cite{5} to me.

The work was supported by the Russian Fund for Basic Research (grant 96-02-16347)
and by INTAS (grant 93-79).

\end{document}